\documentstyle[aps,multicol,epsf]{revtex}

\begin{document}
\draft

\title{Structure of Growing Networks: Exact Solution of the Barab\'{a}si--Albert's Model}

\author{S.N. Dorogovtsev$^{1, 2, \ast}$, J.F.F. Mendes$^{1,\dagger}$, and 
A.N. Samukhin$^{2, \ddag}$}

\address{
$^{1}$ Departamento de F\'\i sica and Centro de F\'\i sica do Porto, Faculdade 
de Ci\^encias, 
Universidade do Porto\\
Rua do Campo Alegre 687, 4169-007 Porto -- Portugal\\
$^{2}$ A.F. Ioffe Physico-Technical Institute, 194021 St. Petersburg, Russia 
}

\maketitle

\begin{abstract}
We generalize the Barab\'{a}si--Albert's model of growing networks 
accounting for
initial properties of sites and find exactly the distribution of connectivities of the network 
$P(q)$ and the averaged connectivity $\overline{q}(s,t)$ of a site $s$ in the instant $t$ 
(one site is added per unit of time). 
At long times $P(q) \sim q^{-\gamma}$ at $q \to \infty$ and $\overline{q}(s,t) \sim (s/t)^{-\beta}$ at $s/t \to 0$, where the exponent $\gamma$ varies from $2$ to $\infty$ depending on the initial attractiveness of sites.  
We show that the relation $\beta(\gamma-1)=1$ between the exponents is universal.  
\end{abstract}


\begin{multicols}{2}

\narrowtext


Recently opened intriguing scaling properties of a variety of growing networks 
(WWW, neural, social, citations of scientific papers, etc.,
see 
\cite{ws98,ls98,red98,hppj98,ajb99,hub,ba99,baj99,baam99,new99,bw00,asbs00,dmj00} and references therein) challenge us to find general reasons of such behavior. In fact, the studying of networks has a long story \cite{erd,bol,par86,dp86,ds87,mon} but only now we become feel ourselves as subjects inside of the world of developing networks and start appreciate significance of the evolving network problem \cite{ajb99,hub,wdj99}.

Mostly, the interest is concentrated at the distribution of shortest pathes between the different sites of a network \cite{ws98} and at the distribution of the number of connections with a site $P(k)$ \cite{ls98,red98,hppj98,ajb99,hub}. The second quantity is obviously simpler to obtain but even for it, in the case of the networks with scaling behavior, no exact results are known.

A clear model of a growing networks was proposed by Barab\'{a}si and Albert \cite{ba99,baj99}. 
The model presents the most simple mechanism of self-organization of a growing network into a free-scale structure. At each time step a new site is added 
and is connected with old sites of the network through $m$ new links. 
The probability of an old site to get a new link is proportional to the total number of connections with this site. 
It was found in \cite{ba99,baj99} that $P(k) \sim k^{-\gamma}$ 
at long times and large connectivities, where $\gamma=3$ (the approach that seemed at first sight be approximate -- ``mean field'' -- was used). This value is close to that one observed in the network of citations \cite{red98}, but other examples show different values of $\gamma$. Introduction of the aging of sites changes $\gamma$ \cite{dm00} and may even break the scaling behavior \cite{asbs00,dm00}. 

In the present letter, we generalize the Barab\'{a}si--Albert's model in natural way and solve it exactly, i. e. we find the exact form of the distribution of connectivities and some other related quantities of the network. We show also that the appearing scaling relations are valid not only for this particular model but for a wide class of growing networks. 

Let us formulate the model. At each time step the new site with $m$ connections with some old sites appears.  Hence, in each instant $t=1,2,\ldots$, the network consists of $t$ sites connected by $mt$ directed links. 
Note that we consider not a total number of $k_s$ connections with a site but $q_s$, 
the number of incoming links, where $s$ is the number of a site. 
Therefore, for brevity, we call it the connectivity of a site but not $k_s$ as usually. One introduces the following rule to distribute $m$ new links among $t$ old sites in the instant $t+1$. Let the probability 
that a new link points to a given old site $s$ be proportional to the following characteristic of the site:

\begin{equation}
A_s = A^{(0)} + q_s
\,,  
\label{1}
\end{equation}
that may be named its {\em attractiveness}. All sites are borned with some initial attractiveness $A^{(0)}$ but afterwards it increases because of the $q_s$ term. 
In the particular case of the Barab\'{a}si--Albert's model, $A^{(0)} = m$. 

In fact we consider the following general problem. In each instant, $m$ new particles 
(i. e. incoming links) have be distributed between an {\em increasing} number (by one per time step) of boxes (i. e. sites) acording to the introduced rule. One has to emphasize, that in our formulation, there is no difference, from which particular site -- old or new or even from outside of the network -- these new links come. 
Indeed, we are interested only in the statistics of incoming links. 
Therefore, new links may also appear between old sides.

It is convenient to assume that in the initial instant $t=1$ we have only one site with the connectivity $m$. Then the total attractiveness of the network in the instant $t$ is $A_\Sigma = (m + A^{(0)})t = (1+a)m$, where $a \equiv A^{(0)}/m$. The resulting behavior at long times is independent of the initial condition. 

Let us derive the equation for the distribution $p(q,s,t)$ of connectivities $q$ at the site $s$ 
in the instant $t$. The probability that a new link is connected with the site $s$ is 
$W_s = A_s/A_\Sigma$. The probability for the
site to receive exactly $l$ new links of $m$ injected is ${\cal P}_s^{(ml)}={\textstyle {m \choose l}}
W_s^l\left( 1-W_s\right) ^{m-l}$. Note that  
we allow multiple links unlike the original Barab\'{a}si--Albert's model, i. e. the connectivity of a given site may increase by more than one in the same instant. That is inessential at long times, when
probability to receive more than one link of $m$ introduced simultaneously is
vanishingly low. Hence, the connectivity distribution of a site obeys the
following master equation:
\end{multicols}
\widetext
\noindent\rule{20.5pc}{0.1mm}\rule{0.1mm}{1.5mm}\hfill

\begin{equation}
p\left(q, s,t+1\right) =\sum_{l=0}^m{\cal P}_s ^{(ml)}p\left(q-l, s,t\right)
=\sum_{l=0}^m {m \choose l}
\left[ \frac{q-l+am}{\left( 1+a\right) mt}\right] ^l\left[ 1-\frac{q-l+am}{%
\left( 1+a\right) mt}\right] ^{m-l}p\left(q-l, s,t\right) 
\,.  
\label{2}
\end{equation}
Eq. (\ref{2}) is supplied with the initial condition $p\left(q, s,s\right) =\delta
\left( q\right) $, which means that sites are borned with zero connectivity (i. e. without incoming links in our definition). 

The distribution function of connectivities in the all network is

\begin{equation}
P \left( q,t\right) =\frac 1t\sum_{u=1}^tp\left( q,u,t\right) 
\,.
\label{4}
\end{equation}
Summing up Eq. (\ref{2}) over $s$ from $1$ to $t$, one gets

\begin{equation}
\left( t+1\right) P \left(q, t+1\right) -p\left(q, t+1,t+1\right) =
-\left(
t-\frac{q+am}{1+a}\right) P \left( q,t\right) 
+\frac{q-1+am}{1+a}P
\left(q-1,t\right) +O\left( \frac P t\right) 
\,.
\label{n}
\end{equation}
In the long-time limit, after the transition to continuous-time approximation,
we obtain

\begin{equation}
\left( 1+a\right) t \frac{\partial P}{\partial t}\left(q, t\right) +\left( 1+a\right) P \left(
q,t\right) +\left( q+am\right) P \left( q,t\right) -\left( q-1+am\right)
P \left(q-1,t\right) =\left( 1+a\right) \delta \left( q\right) 
\, .
\label{4a}
\end{equation}
Finally, assuming that the limit, $P \left( q\right) = P \left(q,t\to \infty \right)$, 
exists, we get the following equation for the
stationary connectivity distribution:

\begin{equation}
\left( 1+a\right) P \left( q\right) +\left( q+ma\right) P \left(
q\right) -\left( q-1+ma\right) P \left( q-1\right) =\left( 1+a\right)
\delta \left( q\right) 
\, .  
\label{5}
\end{equation}
\hfill\rule[-1.5mm]{0.1mm}{1.5mm}\rule{20.5pc}{0.1mm}
\begin{multicols}{2}
\narrowtext

To solve Eq.(\ref{5}) one may use Z-transform of the distribution function:

\begin{equation}
\Phi \left( z\right) =\sum_{q=0}^\infty P \left( q\right) z^q
\,.
\label{5a}
\end{equation}
Then one gets from Eq.(\ref{5})

\begin{equation}
z\left( 1-z\right) \frac{d\Phi }{dz}+ma\left( 1-z\right) \Phi +\left(
1+a\right) \Phi =1+a
\,.  
\label{5b}
\end{equation}
\end{multicols}
\widetext
\noindent\rule{20.5pc}{0.1mm}\rule{0.1mm}{1.5mm}\hfill
The solution of Eq. (\ref{5b}) that is analytic at $z=0$ has the following form:

\begin{eqnarray}
\Phi \left( z\right)  =\left( 1+a\right) z^{-1-\left( m+1\right) a}\left(
1-z\right) ^{1+a}\int\limits_0^zdx\,\frac{x^{\left( m+1\right) a}}{\left(
1-x\right) ^{2+a}} =  \phantom{WWWWWWWWWWWWWWWWWWWWW}
\nonumber 
\\
\frac{1+a}{1+\left( m+1\right) a}\left( 1-z\right) ^{1+a}\,\! _2F_1\!\left[ 1+\left(
m+1\right) a,2+a;2+\left( m+1\right) a;z\right] = 
\frac{1+a}{1+\left( m+1\right) a}\,_2F_1\!\left[ 1,ma;2+\left( m+1\right) a;z\right] 
,  
\label{5c}
\end{eqnarray}
\hfill\rule[-1.5mm]{0.1mm}{1.5mm}\rule{20.5pc}{0.1mm}
\begin{multicols}{2}
\narrowtext
where $_2F_1[\ ]$ is the hypergeometric function. Using its 
expansion \cite{be53} in $z$, we obtain, comparing with Eq.(\ref{5a}),

\begin{equation}
P \left( q\right) =\left( 1+a\right) \frac{\Gamma \left[ \left( m+1\right)
a+1\right] }{\Gamma \left( ma\right) }\frac{\Gamma \left( q+ma\right) }{%
\Gamma \left[ q+2+\left( m+1\right) a\right] }
\, ,  
\label{6}
\end{equation}
that is our main result (see Fig. 1).
In particular, when $a=1$, that corresponds to the case $A_s = m+q_s = k_s$, studied in \cite
{ba99,baj99}, we get exactly

\begin{equation}
P \left( q\right) =\frac{2m\left( m+1\right) }{\left( q+m\right) \left(
q+m+1\right) \left( q+m+2\right) }
\,.  
\label{6a}
\end{equation}
This expression in the limit $q \to \infty$ approaches the corresponding result of \cite{ba99,baj99} 
obtained in the frames of an approximate scheme, 
but the prefactors are different. In fact, the ``mean field'' approximation, used in \cite
{baj99}, is equivalent to continuous-$q$ approximation in our discrete-difference equations.
Indeed, if we replace the finite difference with a derivative over $q$, 
we get the expression obtained in \cite{ba99,baj99}.

At $ma+q\gg 1$, the distribution function (\ref{6}) takes the form: 

\begin{equation}
P \left( q\right) \cong \left( 1+a\right) \frac{\Gamma \left[ \left(
m+1\right) a+1\right] }{\Gamma \left( ma\right) }\left( q+ma\right)
^{-(2+a)}
\,.  
\label{7}
\end{equation}
Therefore, we find the scaling exponent $\gamma $ of the distribution function:

\begin{equation}
\gamma =2+a=2+A^{(0)}/m
\,,  
\label{7a}
\end{equation}
where $A^{(0)}$ is the initial attractiveness of a site.

Let us find the distribution function $p(q,s,t)$ of connectivities at the site $s$.
At long times, $t\gg 1$, keeping only two leading terms in $1/t$ in Eq. (\ref{2}), one gets 
\end{multicols}
\widetext
\noindent\rule{20.5pc}{0.1mm}\rule{0.1mm}{1.5mm}\hfill

\begin{equation}
p\left(q, s,t+1\right) =\left[ 1-\frac{q+am}{\left( 1+a\right) t}\right]
p\left(q, s,t\right) +\frac{q-1+am}{\left( 1+a\right) t}p\left(
q-1, s,t\right) +O\left( \frac p{t^2}\right)   
\, .
\label{2a}
\end{equation}
Assuming the scale of time variation to be much larger then $1$,
we can replace the finite $t$-difference with a derivative:

\begin{equation}
\left( 1+a\right) t\frac{\partial p}{\partial t}\left(q, s,t\right) =\left( q-1+am\right) p\left(
q-1,s,t\right) -\left( q+am\right) p\left(q, s,t\right) 
\,.  
\label{3}
\end{equation}
\hfill\rule[-1.5mm]{0.1mm}{1.5mm}\rule{20.5pc}{0.1mm}
\begin{multicols}{2}
\narrowtext
Finally, using Z-transform in the similar way as before, we obtain the solution of Eq. (\ref{3}), i. e. the connectivity distribution of individual sites:

\begin{equation}
p\left(q, s,t\right) = \frac{\Gamma\left( am+q\right) }{\Gamma \left( am\right) q!}
\left( \frac st\right)^{am/(1+a)}
\left[ 1-\left( \frac st\right) ^{1/(1+a)}\right] ^q
\,.  
\label{9}
\end{equation}
Hence, this distribution has an exponentional tail. 
Now one may get also the expression for the average connectivity of a given site:

\begin{equation}
\overline{q}\left( s,t\right)
=\sum_{q=0}^\infty q \, p\left(q, s,t\right) =a\,m\left[ \left( \frac st\right) ^{%
-1/(1+a)}-1\right] \,.  
\label{10}
\end{equation}
Thus, at a fixed time $t$ the average connectivity of an old site $s\ll t$ depends upon its age as $\sim s^{-\beta }$, where the exponent $\beta =1/\left( 1+a\right) $. Therefore, we have the following relation between the exponents of the considered network:

\begin{equation}
\beta \left( \gamma -1\right) = 1
\,,  
\label{11}
\end{equation}
that was obtained in \cite{dm00} with some particular form of $p(q,s,t)$.

We can show that Eq. (\ref{11}) is universal
and may be obtained from the most general suggestions. In fact, we assume only that the averaged connectivity $\overline{q}(s,t)$ and the 
distribution of connectivities $P(q)$ show scaling behavior. 
Then, in the scaling region, the quantity of interest, i. e. the probability $p(q,s,t)$, has to be of the following form: $p(q,s,t)=(s/t)^{\Delta_1}f(q(s/t)^{\Delta_2})$. Here, $\Delta_1 = \Delta_2$ because of the normalization condition for $p(q,s,t)$ at a fixed $s$, 
$\sum_{q=0}^\infty  \, p\left(q, s,t\right) = 1$. Then, the relation $\overline{q}(s,t) \propto (s/t)^{-\beta}$ leads to $\Delta_1 = \Delta_2 = \beta$ (we use the definition (\ref{10})), and finally, inserting $p(q,s,t)$ in such a form into the relation $P(q) \propto q^{-\gamma}$ at large $q$ and $t$, one gets the relation (\ref{11}).  

The particular form of the scaling function $f(\xi), \xi \equiv q (s/t)^{-\beta}$ depends on the specific model of the growing network. In the case under consideration, it follows from Eq. (\ref{9}) that 

\begin{equation}
f(\xi) = 
\frac{1}{\Gamma(am)}\, 
\xi^{am-1}\exp(-\xi)
\label{23}
\, 
\end{equation}  
at   $s/t \to 0, q \to \infty$ and the fixed $q(s/t)^\beta$.
(We used here that $\Gamma(am+q)/q! \to q^{am-1}$ at $q\to\infty$.)

In the limit of zero initial attractivity of sites ($a=0$) all new sites connect only with the first one, since all other sites have no chance to get a new link. In this case, Eqs. (\ref{7a}) and (\ref{11}) give
$\gamma=2$ and $\beta=1$. For $a=1$, i. e. for the Barab\'{a}si--Albert's model, $\gamma=3$ and $\beta=1/2$ \cite{ba99,baj99}. Finally, when $a\to \infty$, i. e. all sites have equal attractivity all time, and the scaling breaks, one sees from  Eqs. (\ref{7a}) and (\ref{11}) that $\gamma\to\infty$ and $\beta\to 0$. 
Note that the ranges of variation of $\gamma$, $(2,\infty)$, and $\beta$, $(1,0)$,  
are the same as for the network with aging \cite{dm00}.

We see that the ``approximate'' approach of \cite{ba99,baj99} gives the proper values for the critical exponents (see also \cite{dm00} for the network with aging). 
Is that only an occasional coincidence? In fact, in these papers, some particular form of $p(q,s,t)$ was used (i) to derive the equation for the averaged connectivity and, therefore, for $\beta$, and (ii) to obtain relation between $\gamma$ and $\beta$. One can noticed, however, that (i) the equation for the averaged connectivity may be obtained for arbitrary $p(q,s,t)$ and (ii) the relation between the scaling exponents is universal, as we have shown above. Thus, the analytical results for the scaling exponents obtained in \cite{ba99,baj99,dm00} turn to be exact.

A two-parameter fitting was proposed in \cite{ta99} to describe the observed distribution of citations of scientific papers \cite{red98}. 
One sees, that the connectivity distribution of the considered growing networks is of the quite different form. It seems, that the difference occurs because we study the {\em growing} structure unlike the approach \cite{ta99} although the question is open. 

In conclusion, we have found the exact form of the connectivity distribution and the averaged connectivity of sites of the natural generalization for the Barab\'{a}si--Albert's model. 
The considered growing network is self-organized into the free-scale structure. The input flow of the 
new links is distributed between the increasing number of sites. In fact we have considered the new version of a sandpile problem.
The scaling exponents are determined by the
value of the initial attractiveness ascribed to every new site. Depending on this quantity, the scaling exponent $\gamma$ of the connectivity distribution takes values from $2$ to $\infty$. We have shown that the relation between the scaling exponents $\gamma$ and $\beta$ is valid for a wide class of growing networks. 
Our simple solution of this problem lets hope to obtain other exact results for growing networks.

The following question remains open. We have found that the exponent $\gamma$ varies between $2$ and $\infty$. Why is $\gamma$ only 
in the range of
$2$--$3$ in real networks \cite{ba99,asbs00}?
\\

SND thanks PRAXIS XXI (Portugal) for a research grant PRAXIS XXI/BCC/16418/98. JFFM 
was partially supported by the project PRAXIS/2/2.1/FIS/299/94. We also thank E.J.S. Lage 
for reading the manuscript and A.V. Goltsev and Yu.G. Pogorelov
 for many useful discussions.\\
$^{\ast}$      Electronic address: sdorogov@fc.up.pt\\
$^{\dagger}$   Electronic address: jfmendes@fc.up.pt\\
$^{\ddag}$     Electronic address: alnis@samaln.ioffe.rssi.ru

\begin{figure}
\epsfxsize=72mm
\epsffile{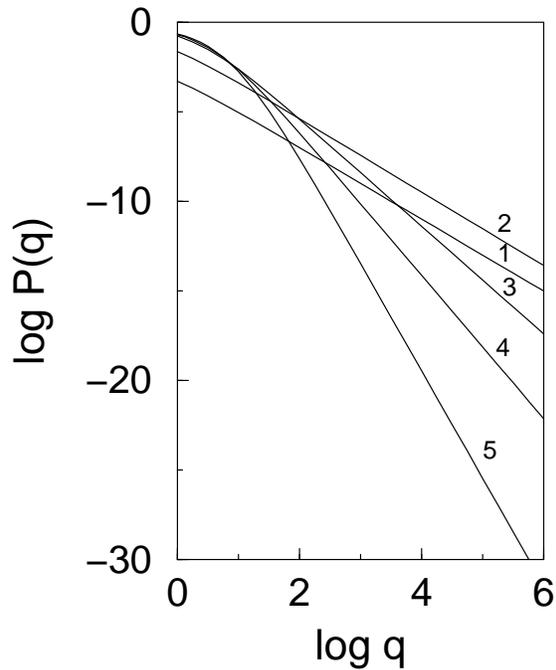}
\caption{
Log-log plot of the distribution of incoming links of sites. $m=1$. 1) $a=0.001$, 
2) $a=0.05$, 3) $a=1.0$ (the Barab\'{a}si--Albert's model), 4) $a=2.0$, 5) $a=4.0$.
}
\label{f1}
\end{figure}

\end{multicols}

\end{document}